\documentclass[referee,usenatbib,usegraphicx]{mn2e}

\title[Emission mechanism of optical pulsations]{Maser mechanism of optical pulsations from anomalous X-ray pulsar
4U0142+61}
\author[Y. Lu, and S.N. Zhang]{Y. Lu $^{1}$\thanks{E-mail:
ly@bao.ac.cn}, and S.N. Zhang $^{2}$
\\
$^{1}$National Astronomical Observatories, Chinese Academy of
Sciences, Beijing, 100012, China\\
$^2$ Physics Department and Center for Astrophysics, Tsinghua
University, Beijing, 100084, China}
\begin{document}

\date{Accepted *******. Received 2004/ 7/30; in original of 2003 December 18}

\pagerange{\pageref{firstpage}--\pageref{lastpage}} \pubyear{2002}

\maketitle

\label{firstpage}

\begin{abstract}
Based on the work of \citet{Luo92}, a maser curvature emission
mechanism in the presence of curvature drift is used to explain
the optical pulsations from anomalous X-ray pulsars (AXPs). The
model comprises a rotating neutron star with a strong surface
magnetic field, i.e., a magnetar. Assuming the space
charge-limited flow acceleration mechanism, in which the strongly
magnetized neutron star induces strong electric fields that pull
the charges from its surface and flow along the open field lines,
the neutron star generates a dense flow of electrons-positions
(relativistic pair plasma) by either two photon pair production or
one photon pair creation resulting from inverse Compton scattering
of the thermal photons above the pulsar polar cap (PC). The motion
of the pair plasma is essentially one-dimensional along the field
lines. We propose that optical pulsations from AXPs are generated
by curvature-drift-induced maser developing in the PC of
magnetars. Pair plasma is considered as an active medium that can
amplify its normal modes. The curvature drift which is energy
dependent is another essential ingredient in allowing negative
absorption (maser action) to occur. For the source of AXP0142+61,
we find that the optical pulsation triggered by curvature drift
maser radiation occurs at the radial distance $R(\nu_M)\sim
4.75\times 10^9$ cm  to the neutron star. The corresponding
curvature maser frequency is about $\nu_M\approx1.39\times
10^{14}$ Hz, and the pulse component from the maser amplification
is about $27\%$. The result is consistent with the observation of
the optical pulsations from the anomalous X-ray pulsar 4U0142+61.
\end{abstract}


\section{Introduction}
Anomalous X-ray pulsars (AXPs) are a class of rare X-ray emitting
pulsars whose energy source have been perplexing for some 20 years
\citep{Aro83,fah81,van95, mer95}. Unlike other x-ray emitting
pulsars, the luminosity of AXPs are orders of magnitude greater
than their rotational spin-down power, and so they require an
additional energy source. One possibility is that AXPs are thought
to be solitary magnetic rotating neutron stars with a magnetic
field stronger than $10^{14}$ G,\citep{Tho96}. This would make
them similar to the soft $\gamma$ ray repeaters
(SGRs)\citep{kou98,Kas04}, but alternative models of binary system
scenario with a very low mass companion that do not require
extreme magnetic fields also exist (for a review see, \cite{Isr02,
Mer02}).

It has been reported that the spectral properties of AXPs are
characterized by the sum of power-law and black-body components
and a small pulse fraction in X-ray band; some AXPs have been
observed with optical emission, which in some cases, i.e., in AXP
4U0142+61, has a larger optical pulse fraction than that in the
X-ray band by five to ten times \citep{Hul00, Ker02, Isr03}. The
optical emission has been related to the accretion disk model
\citep{Cha00} owing to reprocessing of the X-ray irradiation
\citep{Per00}. Assuming that beamed X-rays are emitted at the
neutron star surface, then the rotation of the neutron star leads
to the observed X-ray pulsations. In the accretion disk model
optical pulsations at the X-ray pulsation period can arise from
thermal reprocessing of X-rays illuminating a disk. Another
possible model for 4U0142+61 is a white dwarf with the magnetic
field $B=5\times 10^8$ G and the temperature $T=4\times 10^5$ K,
possibly the result of a double-degenerated merger \citep{Hul00}.
Not only the accretion model and the white dwarf model are
inconsistent with the measurement of optical pulsations, but more
importantly, they are also inconsistent with the observed
optical/IR spectrum which indicates that AXP 4U0142+61 is a
magnetar \citep{Ker02}. The magnetar model, originally proposed by
\cite{Dun92} to explain SGRs, appears to be successful at
interpreting most of the properties of AXPs. A surface dipolar
magnetic field of order $B_{ns}\sim 10^{14}-10^{16}$ G naturally
accounts for the long periods $P$ and high spin-down rates
$\dot{P}$ in the magnetic braking model \citep{kou98}. The
luminous X-ray emission may be explained as the decay of
super-strong magnetic fields \citep{Tho96}. Further evidence has
recently come from the burst activity of two AXPs, as predicted by
the magnetar model \citep{Gav02, Kasp03}. The magnetar model for
AXPs has been spectacularly successful in explaining their most
important phenomenology, including the optical pulsations of AXPs
could arise if the magnetosphere emission of the magnetar is
self-absorbed at optical frequency \citep{Hul00}, and the
magnetars could radiate coherent emission from plasma
instabilities in the infrared and optical bands, as discussed
recently by \cite{Eic02}\,(hereafter EGL02). However there are so
far no detailed models to account for the anomalous behavior noted
for the optical pulsations of AXP\,4U0142+61 \citep{Isr03}.

The possibility of maser mechanism through the inverse population
of particles over energy levels \citep{Chi71, Kap73,Zhe73} has
been widely discussed to address the origin and the main
properties of observed radio emission \citep{Stu71,McC66, Zhe66,
Bes88}. It is maser in a sense that the radiation intensity
exceeds the total spontaneous radiation intensity of all
individual particles in the source, i.e., coherent emission
mechanism acting in astrophysical conditions \citep{Gin75}. In
most models for a pulsar magnetosphere \citep{Gol69, Stu71,
Mic82}, the curvature emission as power mechanism is associated
with such a system. Three theories for coherent curvature emission
have been suggested: coherent curvature emission by bunches
\citep{Stu71,Rud75,Che80}, curvature maser via inhomogeneity due
to curvature of magnetic field lines \citep{Bes88}, and curvature
maser by curvature drift \cite{Zhe79}.

The maser mechanism attributed to curvature drift was discussed in
detail by Luo \& Melrose (1992, hereafter LM92). With an
appropriate choice of the electron energy spectrum, the curvature
absorption coefficient including the curvature drift can become
negative, and the drift motion allows amplification to occur
(LM92). Furthermore, \cite{ Luo95} considers the curvature maser
emission due to field line torsion in pulsar magnetospheres. Such
maser emission was also applied to pulsar radio emission. For both
curvature-drift-induced maser emission (hereafter, CDIME) and
torsion-induced-maser emission (hereafter, TIME), (i) the maser
action relies on the spectral asymmetry due to the dependence of
the drift angular shift on the Lorentz factor $\gamma$, (ii) maser
emission occurs only in a limited angular rage; and (iii) the
Lorentz factor must exceed a threshold \citep{Luo95}.

There is as yet no generally accepted mechanism for explaining why
the optical emission is more strongly pulsed than the soft X-ray
emission. We investigate how the CDIME model works in the magnetar
environment and associate it with the properties of the optical
pulsations from AXP sources. The organization of the paper is as
follows: in Sect.2, we first make a brief review of the necessary
conditions for CDIME proposed by LM92 and discuss the properties
of relativistic pair plasma along the open field lines in
magnetars. In section 3, we discuss the parameters of the region
where optical emission are supposed to be generated and apply the
theoretical framework developed in the previous section for a
detailed calculation of optical pulsations in magnetar sources. It
is shown that the CDIME mechanism makes it possible to explain the
characteristics of optical pulsations in AXPs 4U0142+61. We should
emphasize that the model is based only on the general
considerations concerning the properties of the flux of
relativistic electron-positron plasma flowing in the magnetosphere
of a magnetar. Finally, summary and discussion of the results are
given.

\section{Curvature-drift-induced maser emission and Pair plasma}
\subsection{Necessary conditions for CDIME}
The maser curvature emission triggered by curvature drift has been
investigated by LM92. This mechanism is outlined as following: the
maser emission is possible only when there is beam-type
distribution of outflowing pairs with the Lorentz factor $\gamma$
satisfying a certain value, which is determined by the geometry of
the pulsar magnetosphere. Due to the strong magnetic field, the
electrons and positrons lose all their perpendicular energy
rapidly to synchrotron radiation and fall to the lowest Landau
levels. Therefore the particle motion in a pulsar magnetosphere is
essentially one-dimensional. In the one-dimensional approximation,
for maser action to occur, both the following conditions must be
satisfied:
\begin{eqnarray}
&& df(\gamma)/d\gamma>0\,\,\,\,,\\
&& d\eta/d\gamma<0\,\,.
\end{eqnarray}
where $f(\gamma)$ is the particle distribution function normalized
to unity, and $\eta$ is the spectral power of a single particle,
which is formulated by
\begin{eqnarray}
&& \eta(\omega,\theta,\gamma)=\frac{q^2\omega^2R_B}{6\pi^3c^2}
\{(\theta-\theta_d)^2[\xi^{-1}K_{1/3}(y)]^2+[\xi^{-2}K_{2/3}(y)^2]\}\,\,,
\end{eqnarray}
where $\theta$ is a polar angle with $|\theta|<<1$, $\omega$ is
the curvature emission circular frequency,
\begin{eqnarray}
&&\xi=\{2(1-n)+n[\gamma^{-2}+(\theta-\theta_d)^2]\}^{-1/2}\,\,,\nonumber
\end{eqnarray}
$n$ is the refractive index, and $y$ is defined by
\begin{eqnarray}
&& y=\omega/(3n^{1/2}\omega_R\xi^3)\,\,,\nonumber
\end{eqnarray}
with $\omega_R\approx c/R_B$. $R_B$ is the the radius of curvature
on a field line. Considering a beam of electrons propagation in pure
dipole curved magnetic field, the curvature radius on the
field line can be written as \citep{Lyu99},
\begin{eqnarray}
&& R_B=\frac{4}{3}\frac{\sqrt{R_{ns}R}}{\alpha_*}\,\,\,,
\end{eqnarray}
here $\alpha_*$ is the angle at which a given field line
intersects the neutron star surface. $R$ and $R_{ns}$ is the
radial distance to the star and the surface radius of the star,
respectively. For the case of the last open field, one has
\citep{Lyu99}
\begin{eqnarray}
&& \alpha_*=\sqrt{2\pi R_{ns}/cP}\approx 5.91\times
10^{-3}P_6^{-1/2}\,\,\,,
\end{eqnarray}
where $P$ is the spin period of the star, and $P_6=P/6\,\rm s$.

Putting $\alpha_*$ into Eq.(4), we have
\begin{eqnarray}
&& R_B=2.25\times 10^8P_6^{1/2}\,\,,\\
&& \omega_R(R_B)\approx c/R_B= \frac{4}{3}\times
10^2P_6^{-1/2}\,\,.
\end{eqnarray}
The condition in Eq.(1) requires an effective particle population
inversion which provides free energy to drive the maser emission.
The second condition in Eq.(2) is obtained by partial integration
of the absorption coefficient $\Gamma$, which is written as
\citep{Mel78}
\begin{eqnarray}
&& \Gamma(\omega,\theta)=-\frac{(2\pi c)^3N_0}{2\omega^2mc^2}\int
d\gamma\frac{df(\gamma)}{d\gamma}\eta(\omega,\theta,\gamma)\,\,\,,
\end{eqnarray}
where $N_0$ is the number density of primary relativistic electron
beam ejected from the PC, and is roughly equal to the
Goldreich-Julian density $n_{GJ}$ at the stellar surface. $n_{GJ}$
is defined by \citep{Gol69}
\begin{eqnarray}
&& N_0=n_{GJ}= 1.15\times
10^{12}B_{ns,14}P_{6}^{-1}\,\,\,,\nonumber
\end{eqnarray}
where $B_{ns}$ is the magnetic field strength at the stellar
surface, and $B_{ns, 14}=B_{ns}/10^{14}$ G.

Once the condition in Eqs.(1) and (2) are satisfied, maser action
can occur. It should be noted that maser emission corresponds to
the absorption coefficient Eq.(8) being negative for a certain
range of parameters (e.g. within a certain angular range). In
order to determine in which frequency range the amplification is
important, one needs to estimate the optical depth. According to
LM92, the maximum modulus of optical depth for negative absorption
is estimated to be
\begin{eqnarray}
&&
\tau(a)\approx\frac{1}{8\pi}(\frac{\omega_B}{\gamma\omega})(\frac{W_p}{W_m})\triangle
\theta_0\,\,\,,
\end{eqnarray}
where $\omega_B=|e|B/m_ec$ is the non-relativistic gyro-frequency,
$e$ is the charge of the pair particle, $m_e$ is its mass and $c$
is the speed of light. $W_p=m_ec^2\gamma N_0$ is the particle
energy density, and $W_m=B^2/8\pi$ is the magnetic energy density.
The one-dimensional motion requires that $W_p/W_m\leq 1$.
$\triangle\theta_0$ is defined by
\begin{eqnarray}
&& \triangle \theta_0 = \left\{ \begin{array} {l@{\quad;\quad}l}
 (3\omega_R/\omega)^{1/3} & for \,\,\,\omega \ll \omega_c\,\,,\\
1/\gamma & for\,\,\, \omega \approx \omega_c\,\,,\\
(\omega_c/\omega)^{1/2}\gamma^{-1} & for \,\,\,\omega \gg
\omega_c\,\,,
\end{array}
\right.
\end{eqnarray}
where $\omega_c$ is the characteristic frequency of the curvature
radiation,
\begin{eqnarray}
&& \omega_c=\gamma^3\omega_R\,\,.
\end{eqnarray}
Considering Eq.(10), $\omega_R$, $W_p$ and $W_B$, we rewrite
Eq.(9)
\begin{eqnarray}
&& \tau(a)\approx\left\{ \begin{array} {l@{\quad;\quad}l}
 4.48\times 10^3P^{-1}R_B^{-1/3}\omega^{-4/3} & for \,\,\,\omega \ll \omega_c\,\,,\\
\gamma^{-1}\omega^{-1}P^{-1} & for\,\,\, \omega \approx \omega_c\,\,,\\
1.73\times 10^5\gamma^{1/2} R_B^{-1/2}\omega^{-3/2}P^{-1} & for
\,\,\,\omega \gg \omega_c\,\,,
\end{array}
\right.
\end{eqnarray}
Given the the optical depth $|\tau(\nu_m)|$, one can estimate the
corresponding amplification ratio $\eta$ at maser action occurring
\begin{eqnarray}
&& \eta \equiv\frac{\Delta I_r}{I_r}=\exp(-|\tau(\nu_M)|)\,\,\,,
\end{eqnarray}
where $I_r$ is the intensity of curvature emissions, $\Delta I_r $
is the amplified intensity at maser frequency $\nu_{M}$.
Consequently appreciable amplification requires that the modulus
of the effective optical depth $|\tau(\nu_M)|$ be greater than
unity. According to the conclusions of LM92, the amplification is
important only when the frequencies satisfy $\omega \leq
\gamma^{-1}\omega_B$. This only takes place in the case $\omega\ll
\omega_c$, the corresponding maser frequency $\nu_M$ and the maser
optical depth $|\tau(\nu_M)|$ are, respectively
\begin{eqnarray}
&& \nu_{M}=(2\pi\gamma)^{-1}\omega_B\,\,\,,\\
&&|\tau(\nu_M)|=4.48\times10^3\gamma^{4/3}P^{-1}R_B^{-1/3}\omega_B^{-4/3}\,\,\,.
\end{eqnarray}
To apply the analytic CDIME model described in the above
subsection in the magnetar environment, we need to know the pair
plasma density $n_\pm$ and the Lorentz factor $\gamma$.

\subsection{Pair plasma generation in magnetar environment}
Rotating magnetized neutron stars are unipolar inductors that
generate huge potential drops across the open field line region.
Under certain conditions, a part, or even the total amount, of
this potential will drop across a charge-depleted region (or a
gap) formed in the PC area of the pulsar. Generally, depending on
the boundary conditions at the surface, there are two kinds of
inner accelerator models that are involved to produce the
acceleration of particles and the production of electron-positron
pairs. One is the vacuum (V) type gap if ${\Omega}\cdot {B}<0$
\citep{Rud75}: positive ions are strongly bounded to the surface
so a vacuum gap develops along open field lines above PC, and
pairs are required to provide current flow through the gap, which
can then operate as a stable accelerator. Another is the
space-charge-limited flow (SCLF) type accelerator in polar cap
models \citep{Aro79, Dau96, Har98}: acceleration occurs in the
region of open field near the magnetic poles. On the open field
lines, the neutron star generates a dense flow of
electron-positron pairs penetrated by a highly relativistic
primary electrons. This provides copious secondary pairs flowing
out along the open field lines above the PC of pulsars. As an
underlining PC accelerator model, we employ in this paper the
general relativistic version of a SCLF model developed earlier by
\cite{Mus92} and advanced in a number of important aspects by
Harding \& Muslimov (1998, 2001, 2002), because this kind of SCLF
accelerator can also work in a magnetar
environment\,\citep{Zhaa00}. Although numerical simulations
\citep{Bar01} show that the pair yields drop steeply with
increasing magnetic field, so that in magnetar environments where
photon splitting may effectively suppress pair production at the
stellar surface and consequently the V-type accelerators cannot
develop. However, SCLF gaps could be extremely long and narrow so
that pair formation fronts (PFFs) could be formed at much higher
altitudes above the neutron star surface \citep{Har98, Zha00}.
This means that at the top of SCLF gap when pair production starts
to overcome photon splitting, the pair-production rate also rises
steeply to provide copious pairs \citep{Zhaa00}. Here we assume
that in magnetar environments copious pairs is possible if an SCLF
accelerator is formed. Alternatively, even if photon splitting
could completely suppress one-photon pair production in
superstrong magnetic fields, two-photon pair production more
likely occurs near the threshold, therefore, the magnetar
environment may not be pairless \citep{Zhn01}.

We first investigate the Lorentz factor $\gamma_0$ of a primary
electron accelerated from the PC. For a given distribution of
voltage, the primary particles are accelerated through the
potential drop up to the energies corresponding to the Lorentz
factor $\gamma_0$, whose value is uncertain \citep{Dau82} but is
assumed to be a free parameter here. We follow the acceleration
model of \cite{Har02} and consider the deceleration owing to the
curvature-radiation reaction as \cite{Lyu99} to determine the
ranges of $\gamma_0$
\begin{eqnarray}
&& 3\times 10^6 \le \gamma_0 \le 3 \times 10^{7}\,\,\,\,.
\end{eqnarray}
Pair creation in SCLF model allows an entirely different relation
between charge and current densities $n_{GJ}$ and $J$, the total
number of charge carriers is greatly amplified \citep{Tho02}. And
the pair plasma density $n_{\pm}$ can be scaled roughly by
$n_{GJ}$ at the stellar surface \citep{Bes88, Lyu99},
\begin{eqnarray}
 n_\pm(R_{ns}) &=& \lambda_0 n_{GJ}\,\,\,,\nonumber\\
&=& 1.15\times 10^{12}\lambda_0 B_{ns,14}P_{6}^{-1}\,\,\,,
\end{eqnarray}
where $\lambda_0$ is the multiplicity factor, which varies
theoretically in the range $\lambda_0\sim 10^3-10^6$ \citep{Aro04,
Lyu04, Lyu99, Zha04, Hib01}. According to the relation between the
parameters of the plasma and the beam comes from the energy
argument that the primary particles stop producing pairs when the
energy in the pair plasma becomes equal to the energy in the
primary beam \citep{Lyu99}, we have
\begin{eqnarray}
&& 1.5\times 10^5 <\lambda_0\approx0.05\gamma_0 < 1.5\times 10^6
\,\,\,,
\end{eqnarray}
In the above equation it has been assumed that the initial
density, temperatures and velocities of the plasma components are
equal.


\section{CDIME in Magnetar}
We generalize the analytic expressions for the CDIME model, of the
type described in section 2.1, to a magnetar environment and then
apply it to optical emission of AXPs. The treatment is to assume
that the parameters of CDIME model involved to describe the maser
action in the magnetar environment change with the radial distance
to the neutron star surface.
\subsection{Radius$-$to$-$parameters mapping}
The radial dependence of the parameters is assumed to follow the
dipole geometry of the magnetic field as treated by \cite{Lyu99},
consequently,
\begin{eqnarray}
&& B(r_*)=B_{ns}r_*^{-3}\,\,\,,\nonumber\\
&& \lambda(r_*)=\lambda r_*^{-3} \,\,\,,\nonumber\\
&& \omega_B(r_*)=\omega_B(R_{ns})r_*^{-3}\,\,\,\nonumber\\
&&\gamma(r_*)=\gamma_0 r_*^{-3}\,\,\,,\nonumber\\
&& \omega_R(r_*) =\omega_Rr_*^{-3}\,\,\,,
\end{eqnarray}
 where $r_*=R/R_{ns}$, $R$ is the radial distance to the star,
$R_{ns}=10^6$ cm is the stellar radius. The above relationships in
Eq.(19) are regarded as a `radius to the parameters' mapping.

Because the height of the pair-formation front (PFF), the location
where the first pairs are produced in the SCLF model is larger
compared to the scale of the pair cascade multiplicity growing
\citep{Har02, Lyu99}, so we can assume approximately,
\begin{eqnarray}
&& \lambda(r_*) =\lambda r^{-3}_*= \left\{ \begin{array}
{r@{\quad;\quad}l}
 \lambda_0 & if\,\,\,
r_* \leq r_{*,LS}\\
\lambda_0r_*^{-3} & if\,\,\, r_*> r_{*,LS}\,\,,
\end{array}
\right.
\end{eqnarray}
where $r_{*,LS}=R_{LS}/R_{ns}$, $R_{LS}=Pc/2\pi \simeq 3\times
10^{10}P_6$ cm is the light cylinder radius with the rotational
period of the star $P_6$. And $\omega_R(r_*)$ is assumed to
satisfy
\begin{eqnarray}
&& \omega_R(r_*) =\omega_Rr_*^{-3}= \left\{ \begin{array}
{r@{\quad;\quad}l}
 \omega_R(R_B) & if\,\,\,
r_* \leq r_{*,LS}\\
\omega_R(R_B)r_*^{-3} & if\,\,\, r_*> r_{*,LS}\,\,,
\end{array}
\right.
\end{eqnarray}

Taking into account of the fact that each primary particles
accelerated in the gap generates many secondary electron-positron
pairs \citep{Bes88, Aro83, Bes82, Dau83}, the density of the pair
plasma is substantially higher than the primary beam density.
Consequently, we rewrite Eqs.(9) and (10) as a function of $r_*$
by considering Eqs.(19) to (21) and replacing $N_0$ by $n_\pm$:
\begin{eqnarray}
&& \tau[a(r_*)]\approx\lambda(r_*)P^{-1}\omega^{-1}\triangle
\theta(r_*)\,\,\,,
\end{eqnarray}
where $\triangle\theta(r_*)$
\begin{eqnarray}
&& \triangle \theta(r_*) = \left\{ \begin{array}
{l@{\quad;\quad}l}
 7.36\omega^{-1/3}r_*P_6^{-1/6} & for \,\,\,\omega \ll \omega_c\,\,,\\
\gamma_0^{-1}r_*^3 & for\,\,\, \omega \approx \omega_c\,\,,\\
1.15\times 10^1\gamma_0^{1/2}\omega^{-1/2}r_*^{-1/4}P_6^{-1/4} &
for \,\,\,\omega \gg \omega_c\,\,,
\end{array}
\right.
\end{eqnarray}
Correspondingly, in magnetar environment, the maser frequency
$\nu_M$, $\omega_c$ in Eqs.(11) and (14) are, respectively:
\begin{eqnarray}
\nu_M & =& (2\pi)^{-1}\gamma^{-1}\omega_B=(2\pi)^{-1}\gamma_{0}^{-1}\omega_B(R_{ns})\,\,\,,\nonumber\\
 &=&0.286\times10^{15}\gamma_{0,6}^{-1}B_{ns,14}\,\,\,,\\
  \omega_c &=&\gamma^3\omega_R(r_*)=\gamma_0^3\omega_R(R_B)\,\,\,,\nonumber\\
&=&1.34\times 10^{19}\gamma_{0,6}P_6^{-1/2}\,\,\,,
\end{eqnarray}
where $\gamma_{0,6}=\gamma_0/10^6$.\\
Also, the corresponding optical depth for the maser action in
magnetar is
\begin{eqnarray}
 |\tau(\nu_M)| &=& 1.23\times 10^0\lambda_{0}\gamma_{0}^{4/3}\omega_B^{-4/3}(R_{ns})P_6^{-7/6}r_*^{4}\,\,\,\nonumber\\
& = &5.6\times
10^{0}\lambda_{0,5}\gamma_{0,6}^{4/3}B_{ns,14}^{-4/3}P_6^{-7/6}r_{*,4}^{4}\,\,,
\end{eqnarray}
where $\lambda_{0,5}=\lambda_0/10^5$, and $r_{*,4}=r_*/10^4$

Based on the observed spin period $P$ and spin-down rate $\dot{P}$
of a pulsar, the strength of the surface magnetic field can be
estimated by
\begin{eqnarray}
&& B_{ns,14} = 6.4\times10^{5}(P\dot{P})^{1/2}\,\,\,,\nonumber\\
&&\simeq  1.57(P_6\dot{P}_{-12})^{1/2}\,\,\,\,,
\end{eqnarray}
where $\dot{P}_{-12}=\dot{P}/10^{-12}\,s\,s^{-1}$.\\
One can eliminate $B_{ns,14}$ by substituting Eq.(27) into
Eqs.(24) to (26), consequently,
\begin{eqnarray}
&& \nu_M =0.45\times10^{15}\gamma_{0,6}^{-1}P_6^{1/2}\dot{P}^{1/2}_{-12}\,\,\,,\\
&& \nu_c=\omega_c/2\pi =4.98\times
10^{16}\gamma_{0,6}P_6^{-1}\,\,\,,\\
&& |\tau(\nu_M)|=3.06\times
10^{0}\lambda_{0,5}\gamma_{0,6}^{4/3}P_6^{-11/6}\dot{P}^{-2/3}_{-12}r_{*,4}^4\,\,.
\end{eqnarray}
Although curvature maser emission requires the curvature drift, it
is driven by the free energy in the particle beam. Hence, for
maser emission to occur, the Lorentz factor of the particles must
be larger than a certain value. This minimum energy requirement
may be estimated in terms of LM92,
\begin{eqnarray}
\gamma_0 &>&(\omega_B/\omega_R)_{r_{*,0}}^{1/3}\,\,\,,\nonumber\\
&\approx&  5.12\times10^6P_6^{1/6}\,\,,
\end{eqnarray}
where $r_{*,0}=r_*/10^0=1$. Putting Eq.(31) into Eq.(18), we have
\begin{eqnarray}
&& \lambda_{0,5}=2.56 P_6^{1/6}\,\,.
\end{eqnarray}
Substituting the values of $\gamma_0$ and $\lambda_0$ in the above
equation into Eqs.(28) to (30), we have
\begin{eqnarray}
&& \nu_M =8.79\times10^{13}P_6^{1/3}\dot{P}^{1/2}_{-12}\,\,\,,\\
&& \nu_c=\omega_c/2\pi =2.54\times
10^{17}P_6^{-5/6}\,\,\,,\\
&& |\tau(\nu_M)|=69.12\times
P_6^{-13/9}\dot{P}^{-2/3}_{-12}r_{*,4}^4\,\,.
\end{eqnarray}
With the observed optical amplification determinied by
$|\tau(\nu_M)|$, we obtain the corresponding maser position
$R(\nu_{M})$,
\begin{eqnarray}
&& r_{*,4}(\nu_M)=3.47\times
10^{-1}|\tau(\nu_M)|^{1/4}P_6^{13/36}\dot{P}_{-12}^{1/6}\,\,\,,
\end{eqnarray}
where $r_{*,4}(\nu_M)=r_*(\nu_M)/10^4$, and
$r_*(\nu_M)=R(\nu_M)/R_{ns}$.\\
The CDIME model developed for the magnetar environment is applied
to different AXPs/SGRs sources, as shown in Table 1.


\subsection{Optical pulsation from AXP\,4U0142+61 sources}
Evidence for the optical pulsation has been reported in AXP
sources \citep{Ker02}. AXP\,4U0142+61 is the brightest AXP in
X-rays, has no associated supernova remnant and with its spin-down
timescale of $\sim 10^5\,yr$ \citep{Wil99}. This source is
observed at optical band ranging from $10^{14}$ Hz to $10^{15}$
Hz. The remarkable aspect of the optical light curve is that its
modulation amplitude is very large compared to that of the X-ray
light curves. The pulsed fraction of the optical light is about
$27\%$ of its total optical flux, which is five to ten times
greater than that of soft X-rays. The basic data of all AXPs with
known spin periods and spin-down rate detected with ASCA are
summarized in Table 1.

For the source of AXP 4U142+61, $P_6=1.45$, $\dot{P}_{-12}=1.98$,
and $|\tau(\nu_M)|\approx 1.31$ which is estimated from the pulsed
fraction of $\Delta I_r/I_r\approx 27\%$. We obtain the maser
frequency and the corresponding maser position $r_{*}(\nu_{M})$
from Eqs.(33) and (36), respectively
\begin{eqnarray}
&& r_*(\nu_M)= 4.75\times 10^{3}\,\,,\\
&& \nu_M =1.39\times10^{14}\,\,\rm Hz\,,
\end{eqnarray}
The model prediction shows that the optical pulsations from
AXPs\,4U0142+61 occurs at the radius of $r_*(\nu_M)=4.75\times
10^3$ and its maser frequency is $\nu_M=1.39\times 10^{14}$ Hz.

\cite{Hul01} presented near-infrared and optical observations of
the field of the AXP\,2259+586 taken with the Keck telescope. They
have identified the near-infrared counterpart to AXP\,2259+586 at
$\sim10^{14}$ Hz, but no emission was found at frequencies between
$(2-5)\times10^{14}$ Hz. Associating the model investigated in
this paper to the source of AXP\,2259+586, we find that if
curvature-drift maser action occurs at $r_{*}(\nu_M)=4.75\times
10^3$, from Eqs.(35), we have $|\tau(\nu_M)|\approx 4.61$. The
corresponding maser frequency is about $\nu_{M}=6.45\times
10^{13}$ Hz. Our model further predicts that pulse component of
the optical pulsation is $\sim 0.99\%$. The results show that our
model prediction is consistent with the near-infrared observation
of the source AXP\,2259+586.

In addition, our model predicts that the maser emission should be
detectable at K and J bands, because the maser frequency is
$\nu_{M,13}\simeq 8.79\dot{P}_{-12}^{1/3}P_6^{1/2}$, where
$\nu_{M,13}=\nu_M/10^{13}$ Hz. As a matter of fact, infrared and
near-infrared emissions have also been detected in other two AXPs,
namely, AXP\,1708-4009, AXP1048-5937 \citep{Isr03,Wan02}, but no
pulse components have been detected in these sources so far.

Finally, in comparing the emission properties of HBPs with AXPs in
this discussion, we take into account of two young isolated radio
pulsars, $PSR\,J119-6127$ and $PSR\,J1814-1744$ \citep{Cam00},
which have spin parameters similar to the AXPs and very high
inferred magnetic fields, i.e., high magnetic field pulsars
(HBPs). Table 1 shows that no optical maser should occur in the
sources of HBPs. The possible reason is that although both AXPs
and HBPs are rotating high field neutron stars, they have
different orientations of the magnetic axes with respect to the
rotation axes \citep{Zhaa00}. Consequently, for HBPs, one can
safely ignore the two-photon pair-production mechanism
\citep{Zhn01} that is involved to produce copious pairs for
triggering optical maser actions in AXPs.

\section{Discussion and conclusions}
\subsection{Discussion}
We first need to calculate the optical luminosity of the objects
given the estimated distance. Because the total energy source of
the maser coherent optical emission come from the non-thermal
leptons which include the electrons pulled off from the neutron
star surface and the pairs generated through the secondary cascade
processes, it should therefore be limited by the total spin-down
power of the star. The total luminosity available to convert into
optical emission is the spin down luminosity $L_{sd}$ of the
magnetar\,\citep{Eic02}, which is
\begin{eqnarray}
L_{sd} &=& 4\pi^2IP^{-3}\dot{P}\,\,\,\,,\nonumber\\
&\sim & 1.83\times 10^{32}I_{45}P_6^{-3}\dot{P}_{-12}\,\,\,\, \rm
erg\,\rm s^{-1}\,\,,
\end{eqnarray}
where $I$ is the moment of the star inertia, and
$I_{45}=I/10^{45}\,c\,g\,s$. Observationally, a luminosity of
$10^{33}\,erg\,s^{-1}$ ($\sim 10^{-2}$ of the persistent, pulsed
X-ray flux) in coherent optical emission from AXPs could be
detectable at 2.2 $\mu m$ with imminent technology (EGL02), if the
optical emission is beamed into a solid angle with
$\frac{\delta\Omega}{4\pi}\sim 0.1$, the observed coherent optical
emission $L_{obs}$ from AXPs corresponds to
\begin{eqnarray}
L_{obs} &=& (\frac{\delta\Omega}{4\pi})10^{33}\,\,\,\,,\nonumber\\
&\sim & 10^{32}\,\,\,\,\rm erg\,\rm s^{-1}\,\,,\nonumber
\end{eqnarray}
which is consistent with the model prediction in Eq.(39).

Secondly, we address why the maser process is so powerful that it
can use up essentially all the spin-down energy. One of the most
important differences from magnetars and pulsars is the former
current carries a much larger fraction of the total energy budget
than that of the latter. EGL02 argued that most of the energy
budget passes through the magnetosphere currents, assuming that
the emission is non-thermal and probably inverse-Comptonized in
the magnetosphere \citep{Tho02}. Thus, a considerable fraction of
the long-term magnetic energy dissipation in magnetars could be
used up as coherent electromagnetic emission. The coherent
emission of the pulsar is only a small fraction of the spin-down
power, but it can be a much higher fraction of the power in polar
currents (EGL02).

In addition, a $10^5$ pair multiplicity is invoked in this paper.
The required multiplicity is beyond the predicted values from the
current pulsar/magnetar cascade theories, but not impossible: the
study of the eclipsing behavior system PSR\,J0737-3039 A\&B
requires that the pair multiplicity in the pulsar A is of order of
$10^6$\,\citep{Aro04,Lyu04, Zha04}. A value of multiplicity factor
involved for the Crab is $\lambda=10^6$\,\citep{Ken84, Aro79,
Mus03}. For magnetars, the pair production is potentially
suppressed by the processes such as splitting \citep{Bar01}.
However, we adopt the SCLF model developed by \cite{Har98} and
\cite{Zhab00} in which a pair formation front occurs at a higher
altitude, where the effects of the photon-photon pair production
and the inverse Compton off the thermal photons for high-order
pairs become more important than the photon splitting effect
\citep{Zhab00,Zhn01}.

Finally, we address the low frequency turnovers for curvature
maser emission. Regardless of the details of any particular model
for coherent emission, escaped coherent radiation probably has a
minimum frequency of $\nu_p^`$ in the frame of the outflowing
plasma, which gives it a frequency in the observe frame of
$4\pi\nu_p\gamma^{1/2}$, where $\nu_p$ is defined by
$\nu_p=9\times 10^3n_\pm^{1/2}$ ({EGL02}). Then, the low frequency
turnovers for coherent emission can be inferred as
\begin{eqnarray}
&& \nu_{min} \approx 1.13\times 10^5n_\pm^{1/2}\gamma^{1/2}\,\,\,.
\end{eqnarray}
Combining Eqs.(17),(27),(31),(32) and the assumption
$\gamma=\gamma_or_*^{-3}$, we rewrite $\nu_{min}$ at
$r_*=4.75\times 10^3$
\begin{eqnarray}
&& \nu_{min}\approx 10^{12}P_6^{-1/12}\dot{P}_{-12}^{1/4}\,\,\rm
Hz\,.
\end{eqnarray}
The low cutoff frequency for maser emission of different AXPs/SGRs
sources are listed in Table 2. Table 2 shows that $\nu_{min}$ for
AXPs/SGRs is of the order $\sim 10^{12}\,Hz$, making it plausible
that they are radio-quiet. However emission at wavelengths between
optical and radio bands may not be prohibited, as \cite{Zhn01}
predicted.

\subsection{Conclusions}
In this paper, we focus on some examination of the AXP 4U0142+61
optical emission properties. We have demonstrated that our
treatment allows us to explain the observed optical pulsations in
AXPs. The fundamental differences between the
curvature-drift-induced maser mechanism presented in this paper
and the original one proposed by LM92 are as follows:
\begin{enumerate}
\renewcommand{\theenumi}{(\arabic{enumi})}
\item Our model takes into account the curvature-drift maser
action in a magnetar environment and explains the optical
pulsations of AXPs by arguing that the parameters describing the
influences of pair plasma on the curvature emission are as a
function of the radius from pulsar PCs .
\item Our model
incorporates the SCLF accelerator above the pulsar's PC where
copious pairs can be provided due to steeply rising of the
pair-production rate \citep{Zhn01,Zha00,Lyu99}.
\item We consider
the multiplicity of the pair plasma density at the emission region
of $r_{*}(\nu_M)\sim 4.75\times 10^3 $ from pair cascades, while
LM92 considers only the influence of the pairs from the stellar
surface.

In summary, it is possible to account for the optical pulsations
of AXP 4U0142+61 in terms of curvature-drift-induced maser from
relativistic electron (and positrons) at the radial distance to
the neutron star surface about $R=4.75\times 10^9$ cm. The maser
component is about $27\%$ of the optical light, and the
corresponding optical pulsation frequency is about
$\nu_M=1.39\times 10^{14}$ Hz. We predict that significant
optical/IR pulsations should exist in other five AXP/SGR sources,
but not in AXPs\,2259+586 and HBPs as shown by previous
observations (\textit{see.} Table.2).
\end{enumerate}

\section*{Acknowledgments}
We thank the referee for helpful comments and suggestions on the
manuscript. Prof. K.S. Cheng of Hong Kong University is
acknowledged for originally suggesting this work to us and for
carefully reading the manuscript. We are grateful to L. Zhang for
fruitful discussions as well. This work was supported by the
National Natural Science Foundation of China under grant
no.10273011, the National 973 Project (NKBRSF G19990754), and the
Special Funds for Major State Basic Research Projects.

\newpage
\begin{table*}[]
  \caption[]{ SGR AND AXP TIMING PARAMETERS
   }
  \label{Tab:publ-AXP}
 \begin{tabular}{@{}lllllllc@{}}
  \hline
Sources & $P(s)^{(1)}$ & $\dot{P}_{-12}({\rm s s^{-1}})^{(2)}$ &
    ${\rm References}^{(3)}$ & $\nu_{c,17}(\,\rm Hz)^{(4)}$ & $\nu_{M,14}^{(4)}(\,\rm Hz)$
     & $|\tau(\nu_M)|$ & $\eta$
    \\
  \hline
 SGR 1900+14... & 5.16(7) &60.5    & 1, 2 & 2.88 & 6.50 & $0.283$ & $75.28\%$  \\
 SGR 1806-20... & 7.47(3) &115.5  & 4, 5 & 2.12 & 10.1 & $0.108$ & $89.75\%$   \\
 AXP 1048-5937 & 6.45(1) & 32.6    & 6, 7 & 2.39 & 5.14 & $0.311$ & $73.30\%$ \\
 AXP 1841-045 & 11.77(6) &41.6    & 8,9 & 1.45 & 7.09& $0.112$ & $89.58\%$ \\
 AXP 2259+586 & 6.98(2)  &0.488 &10,11 & 2.25 & 0.65 & $4.613$ & $0.99\%$ \\
 AXP 0142+61 & 8.69(7)  & 1.98  & 12 &1.86 & 1.39 & $1.307$ & $26.98\%$  \\
 AXP 1708-4009 & 11 (6) & 19 & 13,14 & 0.60 & 4.68 & $0.212$ & $81.39\%$ \\
PSR\,J1119-6127  & 0.41 (15) & 4.0 & 15 & 7.18 & $0.72$ & $67.35$ & $\sim 0.0$ \\
PSR\,J1814-1744 & 3.98 (15) &0.74 & 15 & 3.57 & $0.66$ & $7.776$ & $0.04\%$ \\

\hline
\end{tabular}
\begin{enumerate}
\renewcommand{\theenumi}{(\arabic{enumi})}
\item Measured period (1 $\sigma$ error in last digit).

\item  Assumed period derivative (from references).

\item  References.-(1) Hurely et al. 1999; (2) Woods et al. 1999;
(3) Murakami et al. 1999;
  (4) Sonobe et al. 1994; (5) Woods et al. 2000;
  (6) Corbet \& Mihara 1997; (7) Paul et al. 2000
  ; (8) Gotthelf \& Vasisht 1997; (9) Gotthelf, Vasisht, \& Dotani 1999;   (10) Kaspi, Chakrabarty, \& Steinberger 1999;
  (11) Cobet et al. 1995; (12) White et al. 1996; (13) Sugizaki et al. 1997; (14) Israel et al. 1999; (15) Camilo et al. 2000

\item $\nu_{c,17}=\nu_c/10^{17}$; $\nu_{M,14}=\nu_M/10^{14}$
\\

\end{enumerate}
\end{table*}

\begin{table*}[]
  \caption[]{ The conditions of synchrotron maser for the optical pulsation in magnetar
    }
 \label{Tab:publ-AXP}
 \begin{center}
 \begin{tabular}{@{}cccccc@{}}
 \hline
Sources & $\nu_{min,12}^{(1)}$ & $\nu_M <<\nu_c^{(2)}$ & Observation &  Model (pulsed)\\
& Optical/IR & Optical/IR  \\
  \hline
 SGR 1900+14...& 2.83 &yes & no/no & optical   \\
 SGR 1806-20...&3.22 &$yes$ & no/no & optical   \\
 AXP 1048-5937 &2.38 &yes & no/yes  & optical   \\
 AXP 1841-045 &2.40 &yes & no/no & optical   \\
 AXP 2259+586 &0.83  &yes & no/yes & no  \\
 AXP 0142+61 & 1.15 & yes & yes/no & optical   \\
 AXP 1708-4009 &1.98  &yes & no/yes & optical   \\
PSR\,J1119-6127 &1.76  &yes & no/no & no  \\
PSR\,J1814-1744 & 0.96 & yes & no/no & no   \\
\hline
\end{tabular}
\end{center}
\begin{enumerate}
\renewcommand{\theenumi}{(\arabic{enumi})}
\item   $\nu_{min,12}=\nu_{min}/10^{12}\,\rm Hz$\\
\item   Refer to Table 1\\
\end{enumerate}
\end{table*}

\end{document}